\documentclass[final,5p,times,sort&compress]{elsarticle}

\usepackage[utf8]{inputenc}
\usepackage[T1]{fontenc}    
\usepackage{hyperref}       

\usepackage{microtype}      
\usepackage{lmodern}

\usepackage{tikz}
\usepackage{xspace}
\usepackage{color}
\usepackage{url}            
\usepackage{booktabs}       
\usepackage{amsfonts}       
\usepackage{amsmath}        
\usepackage{bm}
\usepackage{physics}
\usepackage{nicefrac}       
\usepackage{graphicx}
\usepackage{subcaption}
\usepackage[normalem]{ulem}

\makeatletter
\def\ps@pprintTitle{%
  \let\@oddhead\@empty
  \let\@evenhead\@empty
  \def\@oddfoot{\reset@font\hfil\thepage\hfil}
  \let\@evenfoot\@oddfoot
}
\makeatother

\def\cD{{\mathcal D}}

\def\cF{{\mathcal F}}

\def\cH{{\mathcal H}}

\def\cL{{\mathcal L}}
\def\cM{{\mathcal M}}

\def\cV{{\mathcal V}}

\def\vd{\mathbf{d}}
\def\RR{\mathbb{R}}

\newcommand{\mr}[1]{\mathrm{#1}}

\DeclareMathOperator*{\E}{\mathbb{E}}

\def \qcname {DeePKS-kit\xspace}

\def \figwidth {\linewidth}

\begin{document}

\begin{frontmatter}

\title{DeePKS-kit: a  package for developing machine learning-based chemically accurate energy and density functional models}

\author[1]{Yixiao Chen}
\author[1]{Linfeng Zhang}\ead{linfeng.zhang.zlf@gmail.com}
\author[2]{Han Wang}\ead{wang_han@iapcm.ac.cn}
\author[1,3]{Weinan E
}
\address[1]{Program in Applied and Computational Mathematics, Princeton University, Princeton, NJ, USA}
\address[2]{Laboratory of Computational Physics, Institute of Applied Physics and Computational Mathematics, Huayuan Road 6, Beijing 100088, People's Republic of China}
\address[3]{Department of Mathematics, Princeton University, Princeton, NJ, USA}

\begin{abstract}
 We introduce \qcname, an open-source software package for developing machine learning based energy and density functional models. 
\qcname is interfaced with PyTorch, an open-source machine learning library,
and PySCF, an {\it ab initio} computational chemistry program that provides  simple and customized tools  for developing quantum chemistry codes.
It supports the DeePHF and DeePKS methods.
In addition to explaining the details in the methodology and the software, we also provide an example of developing a chemically accurate model for water clusters.
\end{abstract}


\end{frontmatter}

{\bf PROGRAM SUMMARY}

\begin{small}
\noindent
{\em Program Title:} DeePKS-kit \\
{\em Developer's repository link:} \\ \url{https://github.com/deepmodeling/deepks-kit} \\
{\em Licensing provisions:} LGPL  \\
{\em Programming language:} Python  \\
{\em Nature of problem:}\\
Modeling the energy and density functional in electronic structure problems with high accuracy by neural network models. Solving electronic ground state energy and charge density using the learned model.  \\
{\em Solution method:} DeePHS and DeePKS methods are implemented, interfaced with PyTorch and PySCF for neural network training and self-consistent field calculations. An iterative learning procedure is included to train the model self-consistently.\\
\end{small}


\section{Introduction}
Conventional computational methods for electronic structure problems  follow a clear-cut hierarchy concerning the trade-off between efficiency and accuracy.
The full configuration interaction (FCI)~\cite{pople1987QCI} method should be sufficiently accurate at the complete basis set limit, but it typically scales exponentially with respect to the number of electrons $N$.
Coupled cluster singles, doubles and perturbative triples (CCSD(T))~\cite{jeziorski1981coupled}, the so-called golden standard of quantum chemistry, has a scaling of $\order{N^7}$.
The cost of  Kohn-Sham (KS) density functional theory (DFT)~\cite{kohn1965self} and the Hartree-Fock (HF) method typically scale as  $\order{N^3}$ and $\order{N^4}$, respectively, and some recently developed xc-functionals has reached great accuracy in specific applications\cite{goerigk2017look}. However, in general, chemical accuracy cannot be  easily achieved for these methods, and the design of xc-functionals can take a lot of efforts.

Recent advances in machine learning is changing the state of affairs. Significant progress has been made by using machine learning methods to represent a wide range of quantities directly as functions of atomic positions and chemical species. An incomplete list includes Refs.~\citenum{behler2007generalized, bartok2010gaussian, rupp2012fast, ramakrishnan2015big, chmiela2017machine, schutt2017schnet, smith2017ani, han2017deep, zhang2018deep, zhang2018end, brockherde2017bypassing, grisafi2018transferable, chandrasekaran2019solving, zepeda2019deep, schutt2019unifying}. These methods generally scale linearly, yet require a large amount of training data that is beyond the current capability of high level methods like CCSD(T). Meanwhile, there have been efforts in parametrizing many body wavefunction and using variational Monte Carlo approach to solve the electronic Schr\"odinger equation directly\cite{han2019solving,hermann2020deep,pfau2020ab}. These methods are generally very accurate and do not need any training label, but the computational cost can be very expensive (although remaining in cubic scaling) due to the need of Monte Carlo sampling. 

Lately, new machine learning models has been developed, that target at achieving chemical accuracy, for a wide variety of atomic and molecular systems, at the cost similar to DFT or HF methods, and requiring fewer training labels. These methods can be roughly divided into two classes.
One class is like the post-HF methods, which use the
ground-state electronic orbitals of a underlying model (HF or DFT) as the input, and output the energy difference between the model and the ground truth.
In this regard, machine learning based methods are used to parameterize the dependence of the energy difference on the input orbitals, following certain physics-based principles.
Representative methods in this class include the MOB-ML method~\cite{welborn2018transferability,cheng2019universal}, the DeePHF method~\cite{chen2020ground}, etc.
The other class of the methods is in the spirit of  DFT,  in which machine learning based methods are used to parameterize the energy functional (of the charge density or Kohn-Sham orbitals) and can be solved to get the ground state energy in a self-consistent way. 
Methods in this class include some earlier attempts\cite{snyder2012finding,bogojeski2019density,lei2019design,liu2017imporving,nagai2020completing} that may not be fully self-consistent, the NeuralXC method~\cite{dick2020machine}, and the DeePKS method~\cite{chen2020deepks}, etc. There are also attempts in using differentiable programming\cite{tamayo2018automatic,li2021kohn,kasim2021learning} to impose self-consistency and improve sample efficiency, at the expense of significant higher computational cost in the training procedure.

With the booming of machine learning based methods for quantum chemistry problems, the community is in urgent need of  codes that can serve as a bridge between machine learning platforms and quantum chemistry softwares, promote the transparency and reproducibility of different results, and better leverage the resultant models for applications.
Developing such a code will not only benefit more potential users, but also avoid unnecessary efforts on reinventing the wheel.
In particular, since the field is at its early stage,  good codes should not only implement certain methods in a user-friendly way, but also provide flexible interfaces for developing new methods or incorporating more functionalities.

In this work, we introduce \qcname, an open-source software package, publicly available at GitHub~\footnote{https://github.com/deepmodeling/deepks-kit} under the LGPL-3.0 License, for developing chemically accurate energy and density functional models. 
\qcname is interfaced with PyTorch\cite{PyTorch2019} in one end, and at the other end, it interfaces with
 PySCF\cite{sun2018pyscf}, an {\it ab initio} computational chemistry program that provides a simple, light-weight, and efficient platform for quantum chemistry code developing and calculation.
\qcname supports the DeePHF and DeePKS methods that were developed by the authors earlier.
Furthermore, it is also designed to provide certain flexibilities for, e.g., modification of the model construction, changing the training scheme, interfacing other quantum chemistry packages, etc.

The rest of the paper is organized as follows.
In Section~\ref{sec:theory}, we introduce the theoretical framework of the
DeePHF and DeePKS methods as well as the notations.
In Section~\ref{sec:software}, we provide a brief introduction on how to use \qcname to train different quantum chemistry models and how to use these models in production calculations.
In Section~\ref{sec:example}, we use training DeePHF and DeePKS models for water clusters as an example to show how to use the package.
Finally, we conclude with some remarks for future directions.

\section{Methodology\label{sec:theory}}

We consider a many-body system with $N$ electrons indexed by $i$ and $M$ clamped ions indexed by $I$.
The ground-state energy of the system can be written as 
\begin{equation}
\begin{aligned}
    E_\mr{tot} &= \min_{\Psi} E_0\qty[\Psi\left(x_1,x_2,\dots,x_N\right)] \\
               &= \min_{\Psi} \ev{T+W+V_\mr{ext}}{\Psi},
     \label{eq:e0tot}
\end{aligned}
\end{equation}
where $\Psi\left(x_1,x_2,\dots,x_N\right)$ is the $N$-electron wavefunction,  $T = -\frac{1}{2} \nabla^2$, $W = \frac{1}{2} \sum_{i,j}\frac{1}{|x_i-x_j|}$ and $V_\mr{ext} = \sum_{I,i} \frac{Z_I}{|X_I - x_i|}$ are the kinetic, electron-electron interaction, and ion-electron interaction operators, respectively. 

Following the (generalized) Kohn-Sham approach, we introduce an auxiliary system that can be represented by a single Slater determinant $\Phi = \frac{1}{\sqrt{N}}\mr{det}\qty[\varphi_i\pqty{x_j}]$, where $\{\varphi_i(x)\}$ is the set of one-particle orbitals. 
We define another energy functional $E[\cdots]$, which takes these one-particle orbitals as input:
\begin{equation}
\begin{aligned}
    E_\mr{tot} &= \min_{\Phi} E [\Phi\left(x_1,x_2,\dots,x_N\right)] \\
               &= \min_{\qty{\varphi_i}, \braket{\varphi_i}{\varphi_j} = \delta_{ij} } E [\Bqty{\varphi_i}].
    \label{eq:etot}
\end{aligned}
\end{equation}
Solving this variational problem with respect to $\qty{\varphi_i}$ under the orthonormality condition  $\braket{\varphi_i}{\varphi_j} = \delta_{ij}$ gives us the celebrated self-consistent field (SCF) equation,
\begin{equation}
    \cH[\Bqty{\varphi_j}] \ket{\varphi_i}
    = \varepsilon_i \ket{\varphi_i} \qfor i = 1 \dots N.
    \label{eq:scf}
\end{equation}
where $\cH = \fdv{E}{\bra{\varphi_i}}$ denotes the effective single particle Hamiltonian that usually consists of kinetic and potential terms. This is a non-linear equation and needs to be solved iteratively. 
The key to the Kohn-Sham DFT methods is to find a good approximation of $E$, so that the ground-state energy and charge density obtained by solving Eq. ~\ref{eq:etot} are close  to those obtained by solving Eq.~\ref{eq:e0tot}.

We divide $E$ into two parts, 
\begin{equation}
    \label{eq:ediff}
    E\qty\big[\Bqty{\varphi_i} | \omega] = E_\mr{base}[\Bqty{\varphi_i}] + E_\delta\qty\big[\Bqty{\varphi_i} | \omega],
\end{equation}
where $E_\mr{base}$ is an energy functional of the baseline method, such as the HF functional or the DFT functional with a certain exchange correlation,
and $E_\delta$ is the correction term, whose parameters $\omega$ will be determined by a supervised learning procedure.

We follow Ref.~\citenum{chen2020ground} to construct $E_\delta$ as a neural network model that takes the ``local density matrix'' as input and satisfies locality and symmetry requirements. 
The ``local density matrix'' is constructed by projecting the density matrix onto a set of atomic basis functions $\Bqty{\alpha^I_{nlm}}$  centered on each atom $I$ and indexed by the radial number $n$, azimuthal number $l$, magnetic (angular) number $m$.
\begin{equation}
    \label{eq:projdm}
    \pqty{\cD^I_{nl}}_{mm'} = \sum_i \braket{\alpha^I_{nlm}}{\varphi_i}\!\braket{\varphi_i}{\alpha^{I}_{nlm'}}.
\end{equation}
For simplicity and locality, we only take the block diagonal part of the full projection. 
In other words, the indices $I$, $n$ and $l$ are taken to be the same on both sides, and only angular indices $m$ and $m'$ differ. 

We take the eigenvalues of those local density matrices to ensure the resulting descriptors are rotational invariant
\begin{equation}
    \label{eq:dmeig}
    \vd^I_{nl} = \mr{EigenVals}_{mm'} \left[\left(\cD^I_{nl}\right)_{mm'}\right],
\end{equation}
where $\mr{EigenVals}_{mm'}$ means that for given other indices, take all possible $m$ and $m'$ values, consider them as a square matrix, and calculate the eigenvalues of it.
Using these descriptors as the direct input of a neural network model, the correction energy $E_\delta$ is given by
\begin{equation}
    \label{eq:edelta}
    E_\delta = \sum_I \cF^\mr{NN} \pqty{\vd^I | \omega},
\end{equation}
where $\cF^\mr{NN}$ is a fully connected neural network, parameterized by $\omega$, containing skip connections\cite{he2016deep}.
$\vd^I$ denotes the flattened descriptors, where different $n$ and $l$ indices have been concatenated into a single vector. Results for short alkanes in Ref.~\citenum{chen2020ground} show that the descriptors are able to distinguish different atomic species, as well as different local chemical environments (such as covalent bonds) around atoms.  Detailed description of the neural network structures can be found in \ref{sec:NN}.

We remark that both DeePHF and DeePKS schemes adopt the same construction for the energy correction, i.e.~Eq.~\ref{eq:edelta}.
Their  difference lies in that DeePHF takes the SCF orbitals of the baseline model, while the DeePKS takes the SCF orbitals of the corrected energy functional Eq.~\ref{eq:ediff}. 
The parameters in the DeePHF scheme are obtained by a standard supervised training process.
At the same time,  
the DeePHF scheme can be turned to a variational model, DeePKS, so that the energy and some electronic information can be extracted self-consistently.
In this context, the correction energy brings an additional potential term $\cV_\delta$ in the single particle Hamiltonian,
\begin{equation}
    \cH = \cH_\mr{base} + \cV_\delta,
\end{equation}
where $\cH_\mr{base}$ is the Hamiltonian corresponding to the base model $E_\mr{base}$, and 
\begin{equation}
    \label{eq:vdelta}
    \cV_\delta = \fdv{E_\delta}{\bra{\varphi_i}} = \sum_{Inlmm'}\pdv{E_\delta}{\left(\cD^I_{nl}\right)_{mm'}}
                              \ketbra{\alpha^I_{nlm}}{\alpha^I_{nlm'}}
\end{equation}
is the correction potential that depends on both orbitals $\Bqty{\varphi_i}$ and NN parameters $\omega$.

Similarly, the forces, defined as the negative gradients of the energy with respect to atomic positions,  can be calculated using the Hellmann-Feynman theorem. 
The procedure leads to an additional term that results from the atomic position dependence of the projection basis $\Bqty{\alpha^I_{nlm}}$,
\begin{equation}
\label{eq:force}
\begin{aligned}
    F \qty\big[\{\varphi_i^*[\omega]\} |\omega] 
    = &\ F_\mr{base} \qty\big[\{\varphi_i^*\}]  \\
    & - \sum_{Inlmm'}\pdv{E_\delta \qty\big[\{\varphi_i^*\} |\omega]}{\left(\cD^I_{nl}\right)_{mm'}} \\
    & \quad \sum_i \ev**{ \pdv{ \left( \ketbra{\alpha^I_{nlm}} \right) }{X} }{\varphi_i^*} 
\end{aligned}
\end{equation}
where $\{\varphi_i^*\}$ denotes the minimizer of the total energy functional and we write out the dependence on the parameters $\omega$ explicitly. 
Note that the force depends directly on NN parameters $\omega$, this allows us to include the force  in the loss function 
for the iterative training procedure.

The training of a DeePKS model requires an additional self-consistent condition, i.e., the prediction of energy has to go through a minimization procedure with respect to the KS orbitals. 
We therefore reformulate the task into a constrained optimization problem,\cite{chen2020deepks}
\begin{equation}
\label{eq:train}
\begin{split}
    \min_\omega \quad & \E_{\mr{data}, \lambda_\rho} \Big[  \left(E_\mr{label} -  E_\mr{model} \qty\big[\{\varphi_i\} | \omega ] \right)^2 \\ 
                      & \quad \  + \lambda_f \left(F_\mr{label} -  F_\mr{model} \qty\big[\{\varphi_i\} | \omega ] \right)^2 \Big] \\
    \mr{s.t.}\quad & \ \exists\, \varepsilon_i \leq \mu, \\
                   & \ \Big(\cH \qty\big[\{\varphi_i\} |\omega] \\
                    & \quad     + \lambda_\rho \cV_\mr{pnt} \qty\big[\rho[\{\varphi_i\}] | \rho_\mr{label}] 
                     - \varepsilon_i \Big) \ket{\varphi_i} = 0, \\
                   & \ \braket{\varphi_i}{\varphi_j} = \delta_{ij} \qfor i,j = 1 \dots N
\end{split}
\end{equation}
where we have written the self-consistent condition as a set of constraints.  $\mu$ denotes the chemical potential. 
We solve Eq.~\ref{eq:train} using the projection method. 
In detail, we first optimize the NN parameters $\omega$ without the constraints using the standard supervised training process. 
Then we project the orbitals $\{\varphi_i^*\}$ back to the subset that satisfies the constraints by solving the SCF equations. 
The procedure is repeated until we reach convergence.

The additional term $\lambda_\rho \cV_\mr{pnt} \qty\big[\rho[\{\varphi_i\}] | \rho_\mr{label}] $ in the constraints is a penalty potential aiming to drive the optimization procedure towards the target density $\rho_\mr{label}$. 
It can be viewed as an alternative form of the loss function using density as labels. 
Since the density does not depend explicitly on the NN parameter $\omega$, it cannot enter a typical loss function. 
We therefore use such a penalty term to utilize the density information in the training process.
When $\lambda_\rho$ is zero, the term vanishes and we recover the typical SCF equation. 
Otherwise, we  have $\cV_\mr{pnt} = 0$ if and only if $\rho[\{\varphi_i\}] = \rho_\mr{label}$. 
We can also make $\lambda_\rho$ a random variable to reduce overfitting, with the expectation taken over $\lambda_\rho$ as well.

In practice, we have to solve the SCF equations with a finite basis set expansion. 
We write $\ket{\varphi_i} = \sum_a c_{ia} \ket{\chi_a}$, where $ \ket{\chi_a}$ denotes a pre-defined finite basis set.
Then the projected density matrix is given by 
\begin{equation}
    \label{eq:projdm_bas}
    \pqty{\cD^I_{nl}}_{mm'} = \sum_{i,a,b} c^*_{ia} c_{ib} \braket{\chi_a}{\alpha^I_{nlm}}\!\braket{\alpha^{I}_{nlm'}}{\chi_b}.
\end{equation}
When solving the SCF equations, we have 
\begin{equation}
    \sum_b H_{ab} c_{ib} = \varepsilon_i \sum_b S_{ab} c_{ib},
\end{equation}
where $H_{ab} = \mel{\chi_a}{\cH}{\chi_b}$, $S_{ab} = \braket{\chi_a}{\chi_b}$ and the correction term from our functional is
\begin{equation}
\begin{aligned}
    \label{eq:vdelta_bas}
    \pqty{V_\delta}_{ab} 
    &= \mel{\chi_a}{\cV_\delta}{\chi_b} \\
    &= \sum_{Inlmm'} \pdv{E_\delta}{\left(\cD^I_{nl}\right)_{mm'}} \bra{\chi_a}\ket{\alpha^I_{nlm}}\bra{\alpha^I_{nlm'}}\ket{\chi_b}.
\end{aligned}
\end{equation}
Similarly, the contribution of our correction term in the force (Eq.~\ref{eq:force}) is given by
\begin{equation}
\begin{aligned}
    \label{eq:force_bas}
    F_\delta 
    = & -\pdv{E_\delta}{X} \\
    = & -\sum_{Inlmm'} \pdv{E_\delta}{\left(\cD^I_{nl}\right)_{mm'}} \\
    & \quad \sum_{i,a,b} c_{ia} c_{ib} \pdv{\bra{\chi_a}\ket{\alpha^I_{nlm}}\bra{\alpha^I_{nlm'}}\ket{\chi_b}}{X},
\end{aligned}
\end{equation}
which is similar to the Pulay force and can give us the derivatives with respect to neural network parameters.
These derivatives are necessary for training the DeePKS model.

We remark that it is a normal supervised learning process for training a DeePHF model, but an iterative learning process is required for training a DeePKS model.
We put more details in the model training and inference processes in the next Section.

\section{Software\label{sec:software}}
\begin{figure}[tb]
    \centering
    \includegraphics[width=0.45\textwidth]{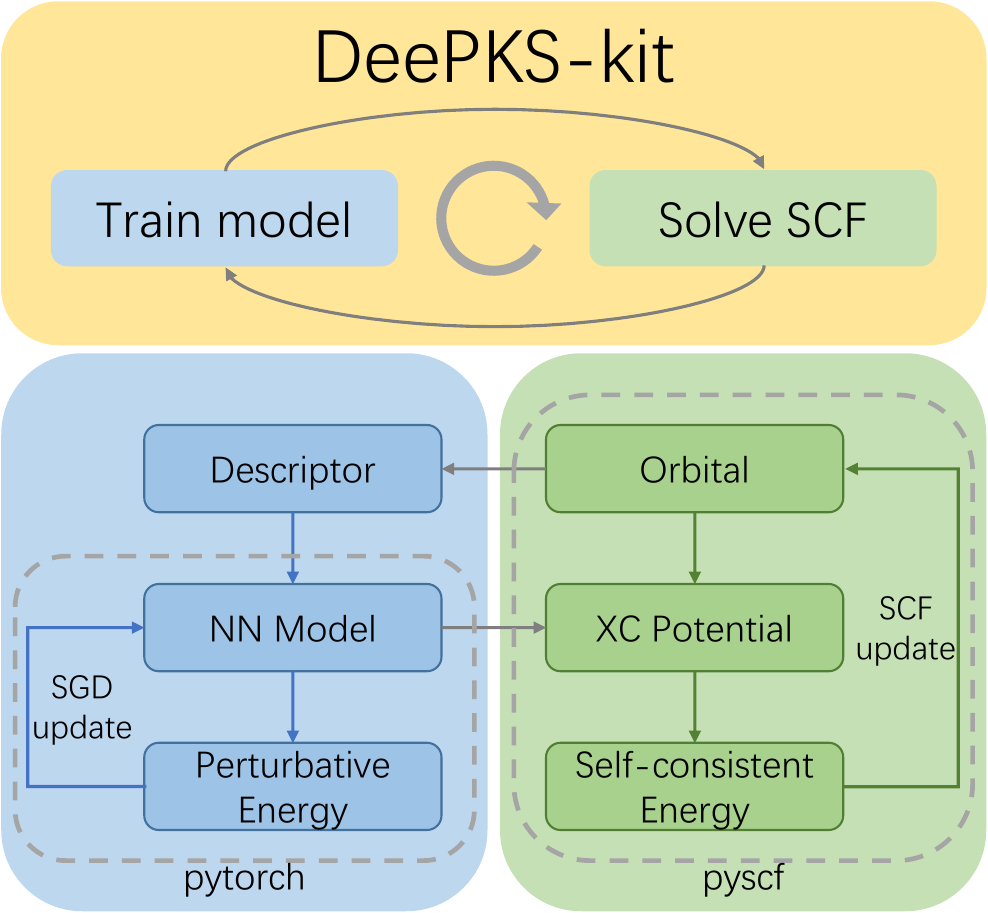}
    \caption{ Schematic plot of the \qcname architecture and the workflow. Upper: main steps of the whole iterative learning procedure. Lower left: training of the neural network (NN) energy functional. Descriptors are calculated from given molecular orbitals and used as inputs of the NN model. The stochastic gradient decent (SGD) training is implemented using the PyTorch library. Lower right: solving generalized Kohn-Sham self-consistent field (SCF) equations. The XC potential is calculated from the trained NN functional. The solver is implemented as a new class of PySCF library.} 
    \label{fig:flow}
\end{figure}

\qcname consists of three major modules that deal with the following tasks: (1) training a (perturbative) neural network (NN) energy functional using pre-calculated descriptors and labels; (2) solving self-consistent field (SCF) equations for given systems using the
 energy functional provided; (3) learning a self-consistent energy functional by iteratively calling tasks (1) and (2). Fig.~\ref{fig:flow} provides a schematic view of the workflow and architecture of \qcname.

\qcname also provides a user-friendly interface, where the above modules together with some auxiliary functions are grouped into a command line tool.
The format reads:
\begin{verbatim}
    deepks CMD ARGS
\end{verbatim}
Currently, \verb|CMD| can be one of the following:
\begin{itemize}
    \item \verb|iterate|, which performs iterative learning by making use of the following commands.
    \item \verb|train|, which trains the energy model and outputs errors during the training steps.
    \item \verb|test|, which tests the energy model on given datasets without considering self consistency.
    \item \verb|scf|, which solves the SCF equation and dumps results and optional intermediate data for training.
    \item \verb|stats|, which examines the results of the SCF calculation and prints out their statistics.
\end{itemize}
\qcname defaults to atomic units (a.u.) with an exception of using Angstrom (\AA) as length unit for \verb|xyz| files.

\subsection{Model training}

The energy functional, as shown in Eq.~\ref{eq:ediff}, is defined to predict the energy difference between a baseline method like Hartree-Fock, or Kohn-Sham DFT, and a more accurate method used for labeling, such as CCSD(T). 
The energy functional is implemented using the PyTorch library as a standard NN model. 
Since the descriptors in Eq.~\ref{eq:dmeig} do not change with the neural network parameters as long as the ground-state wavefunction is given, they are calculated in advance. 
In the current implementation, the atomic basis $\Bqty{\alpha_{nlm}}$ is chosen to be Gaussian type orbital (GTO) functions, so the projection can be carried out analytically. 
We find that the basis needs to be relatively complete, and here we use 108 basis functions per atom and azimuthal indices $l = 0,1,2$. 
One might want to enlarge the set of indices  $l$, but for our testing cases, satisfactory accuracy can already be achieved using the current setup.
The detailed coefficients can be found in the Appendix of Ref.~\citenum{chen2020ground}.
The projection of the density matrix is handled by the library of Gaussian orbital integrals in PySCF. 
The calculation of descriptors can be conducted automatically in the SCF solving part. 
The NN model takes the descriptors $\vd^I_{nl}$ as its input and outputs the ``atomic'' contribution of the correction energy $E_\delta^I = \cF^\mr{NN} \pqty{\Bqty{\vd^I_{nl}}}$, followed by a summation to give the total correction energy $E_\delta$ in Eq.~\ref{eq:edelta}. 

Similarly, the force calculated by the NN model is the difference between the baseline method and the labeling method, namely the second term in Eq.~\ref{eq:force}. 
It can be viewed as an analog of the Pulay force in SCF calculations using GTO basis. 
In the training procedure, similar to energy, the calculation of forces is separated into the parameter-dependent and parameter-free parts, by rewriting the force term using the chain rule,
\begin{equation}
    \pdv{E_\delta}{X} = \sum_{Inl}\pdv{E_\delta}{\vd^I_{nl}} \pdv{\vd^I_{nl}}{X}.
\end{equation}
The parameter-free part $\pdv*{\vd^I_{nl}}{X}$ is pre-calculated and multiplied to the NN-dependent part $\pdv*{E_\delta}{\vd^I_{nl}}$ given by backward propagation to speed up the evaluation.

The training procedure (\verb|train| command) requires the descriptors $\vd^I_{nl}$ to be the input data and the reference correction energies $E_\delta^\mr{label}$ to be the output label. 
If training with forces is enabled, the gradients of the descriptors $\pdv*{\vd^I_{nl}}{X}$ and the reference correction forces $F_\delta^\mr{label}$ are also needed. 
We call the collection of these quantities for a single molecule configuration a \textit{frame}. 
Frames are grouped into Numpy binary files in folders, with names and shapes \verb|dm_eig.npy|: $\qty[n_\mr{frame}, n_\mr{atom}, n_\mr{dest}]$, \verb|l_e_delta.npy|: $\qty[n_\mr{frame}, 1]$, \verb|grad_vx.npy|: $\qty[n_\mr{frame}, n_\mr{atom}, 3, n_\mr{atom}, n_\mr{dest}]$ and \verb|l_f_delta.npy|: $\qty[n_\mr{frame}, n_\mr{atom}, 3]$, where $n_\mr{dest}$ denote the number of projection basis hence descriptors on each atom, and equals to 108 in our tested examples.  
We call each folder a \textit{system}, which also corresponds to the system in the SCF procedure that we will discuss later. 
Frames in the same system must have the same number of atoms, while the type of elements can be different. 
These systems can be prepared manually or generated automatically by the \verb|scf| command described later.

The training of the model follows a standard mini-batch stochastic gradient decent (SGD) approach with the ADAM optimizer\cite{kingma2014adam} provided by PyTorch. 
At each training step, a subset of training data is sampled from one or multiple systems to form a \textit{batch}. 
Frames in the same batch must contain the same number of atoms. 
The training steps are grouped into \textit{epoches}. 
Each epoch corresponds to the number of training steps for which the number of frames sampled is equal to the size of whole training dataset. 
With a user-specified interval of epoches, the square root of averaged loss is output for both training and testing datasets. 
The state of the NN model is also saved at the output step and can be used as a restarting point. 
The saved model file is also used in the SCF procedure.
After  training is finished, \qcname offers a \verb|test| command to examine the model's performance as a DeePHF (non-self-consistent) energy functional.
It takes the model file and system information with the same format in training, and outputs the predicted correction energies for each frame in the system, as well as averaged errors.

\subsection{SCF Solving}

The solver of the SCF equation is implemented as an inherited class from the restricted Kohn-Sham (\verb|RKS|) class in the PySCF library. 
Currently, \qcname only supports restricted calculations in a non-periodic system. 
Necessary tools for supporting unrestricted SCF and periodic boundary conditions will be implemented in our future work. 

For each step in the SCF calculation, our module computes the correction energy $E_\delta$ and the corresponding potential $\cV_\delta$ in Eq.~\ref{eq:vdelta_bas} under a given GTO basis set and adds it to the original potential. 
The calculation of $E_\delta$ is the same as in the training, by calling the PyTorch library to evaluate the NN model, but the descriptors are generated on the fly using the projected density matrices from Eq.~\ref{eq:projdm_bas}. 
The overlap coefficients of GTOs are pre-calculated and saved to avoid duplicated computations. 
The calculation of the potential $\cV_\delta$ follows a similar but reversed approach. 
The gradient with respect to the projected density matrix is computed by backward propagation via PyTorch and then contracted with the overlap coefficients. 
The rest of the SCF calculation, including matrix diagonalization and self-consistent iteration, is handled by PySCF using its existing framework. 

Force computation is also implemented as an extended class of the corresponding \verb|Gradient| class in the PySCF library. 
Once the SCF calculation converges, the additional force term can be computed by following Eq.~\ref{eq:force_bas}, 
using the converged density matrix. 
Adding the additional term to the original force provided by PySCF  gives us the total force acting on an atom.

The \verb|scf| command provided by \qcname is a convenient interface of the module above that handles  loading and saving automatically. 
Similar to the training procedure, this command also accepts \textit{systems} that contains multiple frames grouped into Numpy binary files. 
The data required for each frame is the nuclear charge and position for every atom in that configuration. 
This should be provided in \verb|atom.npy| with shape $\qty[n_\mr{frame}, n_\mr{atom}, 4]$.
The last axis contains four elements corresponding to the nuclear charge and three spacial coordinates respectively. 
For systems for which all frames contain the same element type,  one can provide atomic positions and element type in two separate files: \verb|coord.npy|: $\qty[n_\mr{frame}, n_\mr{atom}, 3]$ and \verb|type.raw|. 
Additionally, \verb|energy.npy|: $\qty[n_\mr{frame}, 1]$ and \verb|force.npy|: $\qty[n_\mr{frame}, n_\mr{atom}, 3]$ can be provided as reference energies and forces to calculate the corresponding labels $E_\delta^\mr{label}$ and $F_\delta^\mr{label}$. 
\qcname takes the name of the folder that contains the aforementioned files as the system's name. 
The current implementation requires all systems to have different names. 
For convenience, the \verb|scf| command also accepts a single \verb|xyz| file as a system that contains one single frame. 
\qcname provides a script that converts a set of \verb|xyz| files into a normal system as well.

The interface takes a list of \textit{fields} to be computed after the SCF calculation, including (but not limited to) the total energy and force, the density matrix, and all the data needed in the training procedure.  
The computed fields will also be grouped into Numpy binary files saved in the folder with the system's name at a specified location. 
The saved folder corresponds to a training system and can be used as the input of the \verb|train| command directly. 
\qcname also provides an auxiliary \verb|stats| command that reads the dumped SCF results and outputs the statistics, including the convergence and averaged errors of the system.

Following Eq.~\ref{eq:train}, the SCF procedure can also accept an additional penalty term that applies on the Hamiltonian in order to use density labels in the iterative learning approach. 
Such penalty is implemented as a hook to the main SCF module that adds an extra potential based on the density difference from the label. 
Currently, $L^2$ and Coulomb norms are supported as the form of the penalty. 
The interface takes an optional key that specifies the form and strength of the penalty. 
To apply the penalty, an additional label file \verb|dm.npy| with shape $\qty[n_\mr{frame}, n_\mr{basis}, n_\mr{basis}]$ is required in the systems.

\subsection{Iterative learning}

The iterative learning procedure is implemented using the following strategy. 
First, we implement a general module that handles the sequential execution of tasks. 
This includes two main parts, 
(1) a scheduler that takes care of the progress of the tasks, creates files and folders, and restarts when necessary;
(2) a dispatcher that submits tasks to, and collects results from, specified computing resources.
Second, to enhance the flexibility of the iteration process, we define a set of task templates that execute of the aforementioned \qcname commands iteratively using user-provided systems and arguments. 
The detailed iteration structure can be easily modified or extended. 
More complicated iterations like active learning procedure can also be implemented with ease. 

The scheduler part consists of tasks and workflows, implemented as corresponding Python classes. 
A task is made up of its command to be executed, its working directory, and the files required from previous calculation. 
Supported execution methods include shell command, Python function, and the dispatcher that handles running on remote machines or clusters. 
A workflow is a group of tasks or sub-workflows that run sequentially. 
When been executed, it will execute the command for each task in the specified order, linking or creating files and folders in the process. 
It will also record the task it has finished execution into a designated file (defaults to \verb|RECORD|), so that if the execution is stopped in the middle, it can restart from its previous location.  

The dispatcher component is adapted from the DP-GEN package\cite{zhang2020dp}. 
It handles the execution of tasks on remote machines or HPC clusters. 
Procedures like file uploading and downloading or job submission and monitoring are all taken care of by the dispatcher. 
Parallel execution of multiple tasks using the same dispatcher is also possible, to ensure that computing resources can be fully utilized. 
Detailed implementation of the dispatcher is explained in Ref.~\citenum{zhang2020dp}. 
Currently, only shell and Slurm systems are supported in \qcname. 
More HPC scheduling systems and cloud machines will be supported in the future. 

With the help of the iteration module described above, we provide a set of predefined task templates that implement the iterative training procedure discussed in Section~\ref{sec:theory}. 
For each iteration, we define four tasks as follows. First, we call the \verb|scf| command to run the SCF calculations using the model trained in the previous iteration on given systems. 
The command is executed through the dispatcher in an embarrassingly parallelized way. 
The SCF procedure will dump results and data needed for training for each system. 
Second, the \verb|stats| command runs directly through Python to check the SCF results and output error statistics. 
Third, the \verb|train| command is executed through the dispatcher that trains the NN model, using the last model as a restarting point, on the data saved by the SCF procedure. 
Last, an additional \verb|test| command is run through Python to show the accuracy of the trained model. The whole workflow consists of multiple number of iterations. For the first iteration where no initial NN model is specified, the SCF procedure will run without any model using a baseline method like HF. 
The first training step will also start from scratch and likely take more epochs. 
The number of iterations, as well as  the parameters used in the SCF and training procedures and their dispatchers, can all be specified by the user through a configuration \verb|YAML| file. 

The \verb|iterate| command reads the configuration file, generates the workflow using the task templates, and executes it. 
If the \verb|RECORD| file exists, the command will try to restart from the latest checkpoint recorded in the file.  The required input systems used as training data are of same format as in the \verb|scf| command. Since this is a learning procedure, the reference energy in \verb|energy.npy| file is required. Force (\verb|force.npy|) and density matrix (\verb|dm.npy|) data are optional and can be specified in the configuration file. 

\section{Example\label{sec:example}}
Here we provide  a detailed example on generating a DeePHF or DeePKS functional for water clusters, and demonstrate its generalizability with tests on water hexamers.
We  use energy and force as labels. An example of training with density labels is also provided in our git repository~\footnote{\url{https://github.com/deepmodeling/deepks-kit/tree/master/examples/water_cluster}}.
We take for example \verb|args.yaml| as the configuration file. The learning procedure can be started by the following command:
\begin{verbatim}
    deepks iterate args.yaml
\end{verbatim}

\textit{System preparation}.
We use randomly generated water monomers, dimers, and trimers as training datasets. 
Each dataset contains 100 near-equilibrium configurations. We also include 50 tetramers as a validation dataset. 
We use energy and force as labels. 
The reference values are given by CCSD calculation with the cc-pVDZ basis. 
The system configurations and corresponding labels are grouped into different folders by the number of atoms, following the convention described in the previous Section. 
The path to the folders can be specified in the configuration file as follows:
\begin{verbatim}
systems_train: 
  - ./systems/train.n[1-3]
systems_test: 
  - ./systems/valid.n4 
\end{verbatim}

\textit{Initialization of a DeePHF model}.
As a first step, we need to train an energy model as the starting point of the iterative learning procedure. 
This consists of two steps. 
First, we solve the systems using the baseline method such as HF or PBE and dump the descriptors needed for training the energy model. 
Second, we conduct the training from scratch using the previously dumped descriptors. 
If there is already an existing model, this step can be skipped, by providing the path of the model to the \verb|init_model| key.
The energy model generated in this step is also a ready-to-use DeePHF model, saved at \path{iter.init/01.train/model.pth}. 
If self-consistency is not needed, the remaining iteration steps can be ignored. 
We do not use force labels when training the DeePHF energy model.

The parameters of the init SCF calculation is specified under the \verb|init_scf| key. 
The same set of parameters is also accepted as a standalone file by the \verb|deepks scf| command when running SCF calculations directly.  
We use cc-pVDZ as the calculation basis. 
The required fields to be dumped are:
\begin{verbatim}
dump_fields: [conv, 
              e_tot, dm_eig, l_e_delta]
\end{verbatim}
where \verb|dm_eig|, \verb|l_e_delta|, \verb|e_tot|, and \verb|conv| denote the descriptors, the labels (reference correction energies), the total energy, and the record of convergence, respectively.
Additional parameters for molecules and SCF calculations can also be provided to \verb|mol_args| and \verb|scf_args| keys, and will be directly passed to the corresponding interfaces in PySCF.

The parameters of the initial training is specified under the \verb|init_train| key. 
Similarly, the parameters can also be passed to the \verb|deepks train| command as a standalone file. 
In \verb|model_args|, we adopted  the neural network model with three hidden layers and 100 neurons per layer, using the GELU activation function\cite{hendrycks2016gaussian} and skip connections\cite{he2016deep}. 
We also scale the output correction energies by a user-adjustable factor of 100, so that it is of order one and easier to learn. 
In \verb|preprocess_args|, the descriptors are set to be preprocessed to have zero mean on the training set. 
A prefitted ridge regression with a penalty strength 10 is also added to the model to speed up the training process. 
The batch size is set to 16 in \verb|data_args|,
and the the total number of training epochs is set to 50000 in \verb|train_args|. 
The learning rate starts at 3e-4 and decays by a factor of 0.96 for every 500 epochs.

\textit{Iterative learning for a DeePKS model}.
For self-consistency, we take the model acquired in the last step and perform several additional iterations of SCF calculation and NN training. 
The number of iterations is set to 10 in the \verb|n_iter| key. 
If it is set to 0, no iteration will be performed, which gives the DeePHF model. 
In the iterative learning procedure, we also include forces as labels to improve the accuracy.

The SCF parameters are provided in the \verb|scf_input| key, following the same rules as the \verb|init_scf| key. 
In order to use forces as labels, we add additional \verb|grad_vx| for the gradients of descriptors and \verb|l_f_delta| for reference correction forces. \verb|f_tot| is also included for the total force results.
\begin{verbatim}
dump_fields: [conv, 
              e_tot, dm_eig, l_e_delta,
              f_tot, grad_vx, l_f_delta]
\end{verbatim}
Due to the complexity of the neural network functional, we use looser (but still accurate enough) convergence criteria in \verb|scf_args|, with \verb|conv_tol| set to 1e-6.

The training parameters are provided in the \verb|train_input| key, similar to \verb|init_train|. 
However, since we are restarting from the existing model, no \verb|model_args| is needed, and the preprocessing procedure can be turned off. 
In addition, we add \verb|with_force: true| in \verb|data_args| and \verb|force_factor: 1| in \verb|train_args| to enable using forces in training. 
The total number of training epochs is also reduced to 5000. 
The learning rate starts as 1e-4 and decays by a factor of 0.5 for every 1000 steps.

\textit{Machine settings}.
How the SCF and training tasks are executed is specified in \verb|scf_machine| and \verb|train_machine|, respectively. 
Currently, both the initial and the following iterations share the same machine settings. 
In this example, we run our tasks on local computing cluster with Slurm as the job scheduler. 
The platform to run the tasks is specified under the \verb|dispatcher| key, and the computing resources assigned to each task is specified under \verb|resources|. 
The setting of this part differs on every computing platform. 
We provide here our \verb|training_machine| settings as an example:
\begin{verbatim}
dispatcher: 
  context: local
  # use "shell" to run without slurm
  batch: slurm 
  # unnecessary in local context
  remote_profile: null 
resources: 
  # resources are ignored in shell batch
  time_limit: `24:00:00'
  cpus_per_task: 4
  numb_gpu: 1
  mem_limit: 8 # gigabyte
python: "python" # use python in path
\end{verbatim}
where we assign four CPU cores and one GPU to the training task, and set its time limit to  24 hours and memory limit to  8GB. 
The detailed settings available for \verb|dispatcher| and \verb|resources| can be found in the example folder of our git repository, as well as in the document of the DP-GEN software, with a slightly different interface. In the case that the Slurm scheduler is not available, we also provide in the repository an example input file to run the tasks in standard shell environment.

\begin{figure}[tb]
    \centering
    \begin{subfigure}{\figwidth}
    \begin{tikzpicture}
        \node[inner sep=0pt] (fig) at (0,0)
        {\includegraphics[width=\textwidth]{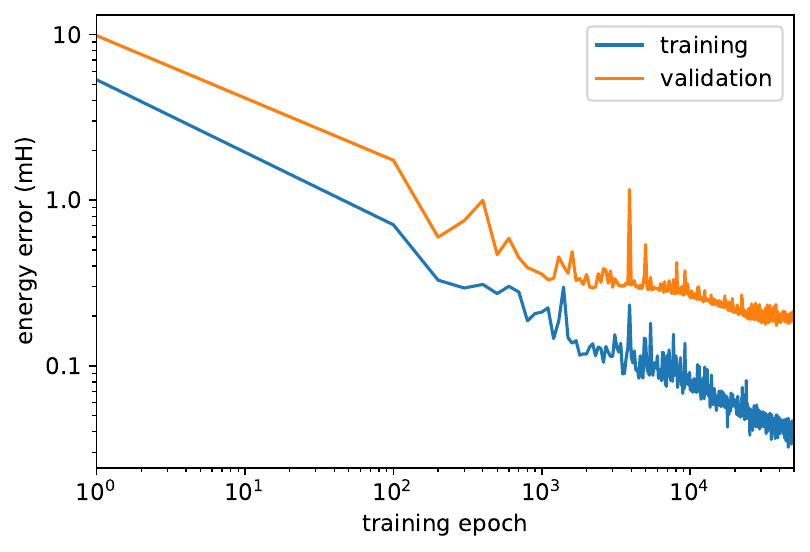}};
        \node[inner sep=0pt, anchor=west] at (fig.north west) {\footnotesize (a)};
    \end{tikzpicture}
    \end{subfigure}
    \begin{subfigure}{\figwidth}
    \begin{tikzpicture}
        \node[inner sep=0pt] (fig) at (0,0)
        {\includegraphics[width=\textwidth]{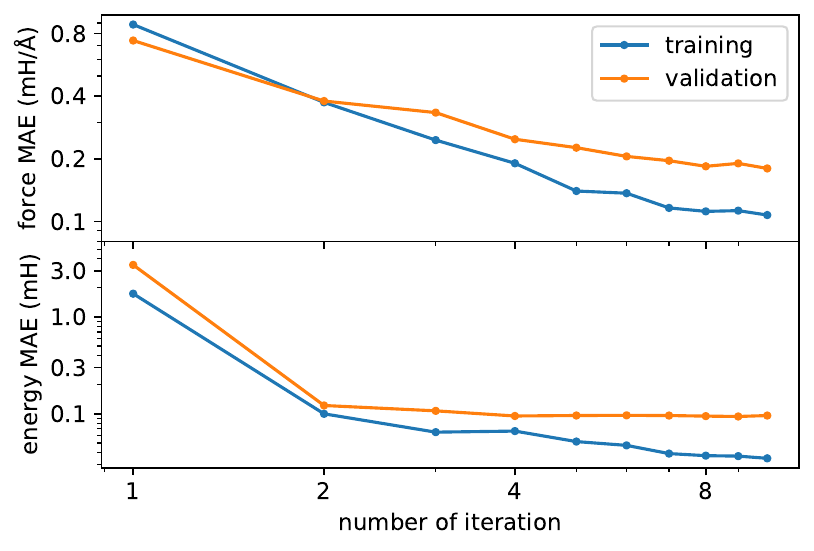}};
        \node[inner sep=0pt, anchor=west] at (fig.north west) {\footnotesize (b)};
    \end{tikzpicture}    
    \end{subfigure}
    \caption{(a) The energy error during the initial training process. (b) The energy and force error during the iterative learning procedure. Both training and validation datasets are evaluated. Axes are plotted in log scale.}
    \label{fig:train}
\end{figure}

\textit{Model testing}.
During each iteration of the learning process, a brief summary of the accuracy of the SCF calculation can be found in \path{iter.xx/00.scf/log.data}. 
Average energy and force (if applicable) errors are shown for both the training and validation dataset. 
The corresponding results of the SCF calculations are stored in \path{iter.xx/00.scf/data_test} and \path{iter.xx/00.scf/data_train}, grouped by training and validation systems. 
The (non-self-consistent) error during each neural network training process can also be found in \path{iter.xx/01.train/log.train}.
We show in Fig.~\ref{fig:train} the energy errors during initial training process and the energy and force errors of SCF calculation in each iteration.
At the end of  the training, all errors are much lower than the level of chemical accuracy.
After  10 iterations, the resulted DeePKS model can be found at \path{iter.09/01.train/model.pth}. The model can be used in either a Python script creating the extended PySCF class, or directly the \verb|deepks scf| command. 

\begin{figure}[tb]
    \centering
    \includegraphics[width=\figwidth]{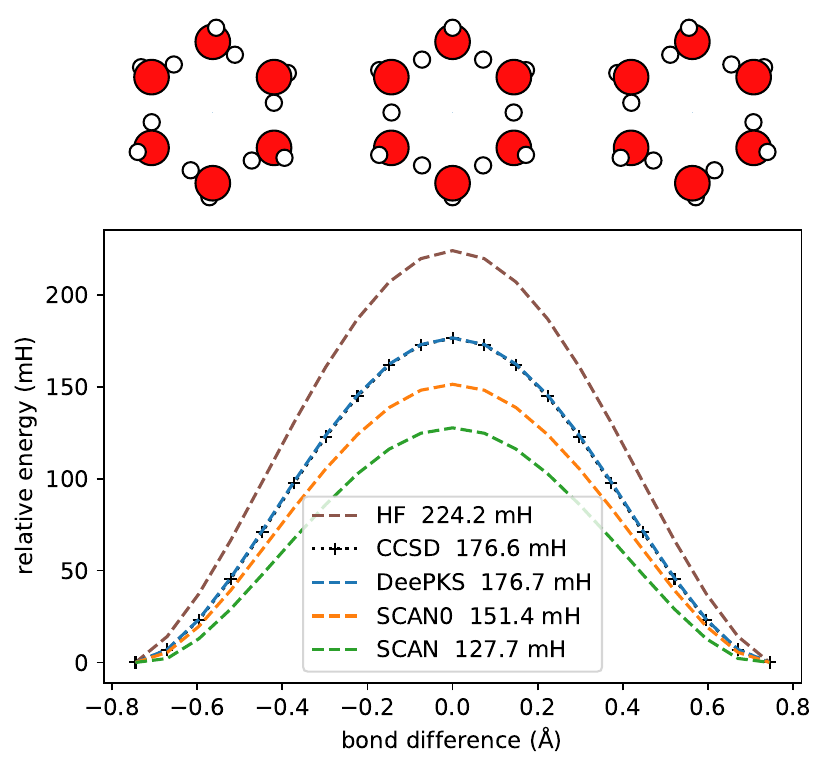}
    \caption{Energy barrier of the simultaneous proton transfer in a water hexamer ring, calculated by different methods. The $x$ coordinate corresponds to the length difference between two OH bonds connecting to the transferred Hydrogen atom. Barrier heights during the transfer are also shown in inset. } 
    \label{fig:ptrans}
\end{figure}

As a testing example, we run the SCF calculation using the learned DeePKS model on the collective proton transfer reaction in a water hexamer ring.
We show the energy barrier of the proton transfer path in Fig.~\ref{fig:ptrans}.
All predicted energies from the DeePKS model fall within the chemical accuracy range of the reference values given by the CCSD calculation. 
We note that none of the training dataset includes dissociated configurations in the proton transfer case. 
Therefore, the DeePKS model trained on up to three water molecules exhibits a fairly good transferability, even in the OH bond breaking regime.

\begin{figure}[tb]
    \centering
    \includegraphics[width=\figwidth]{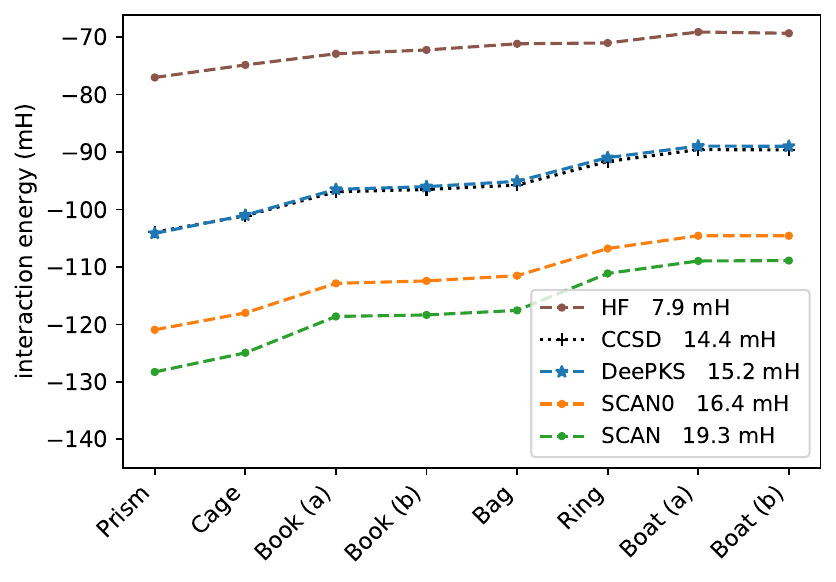}
    \caption{Binding energies of different isomers of water hexamers, calculated by different methods. The values shown inside correspond to the energy difference between conformations with the highest (Boat (a)) and the lowest (Prism) energies.} 
    \label{fig:hexa}
\end{figure}

We also perform another test by calculating the binding energies of different isomers of the water hexamer. 
The binding energy is given by $\Delta E^x = E \pqty\big{\pqty{\mr{H}_2\mr{O}}_6^x} - 6E \pqty\big{\mr{H}_2\mr{O}}$ for a specified isomer $x$. 
The conformations of different water hexamers and the reference monomer are taken from Ref.~\citenum{lambros2020how} with geometries optimized at the MP2 level. 
The results are plotted in Fig.~\ref{fig:hexa}. 
We can see that the DeePKS model gives chemically accurate predictions for all isomers, outperforming the commonly used conventional functionals like SCAN0\cite{sun2015SCAN,hui2016scan}. 
Meanwhile, the relative energies between different isomers are also captured accurately with the error less than 1~mH.


\section{Conclusion\label{sec:conclusion}}
We have introduced the underlying theoretical framework, the details of software implementation, and an example for the readers to understand and use the \qcname package.
More capabilities, such as  unrestricted SCF and periodic boundary conditions, will be implemented in our future work. 
Moreover, we hope that this and subsequent work will help to develop an open-source community that will facilitate a joint and interdisciplinary effort on developing universal, accurate and efficient computational quantum chemistry models.

\section*{Acknowledgement}
The work of Y. C., L. Z. and W. E was supported in part by a gift from iFlytek to Princeton University, the ONR grant N00014-13-1-0338, and the Center Chemistry in Solution and at Interfaces (CSI) funded by the DOE Award DE-SC0019394.
The work of H. W. is supported by the National Science Foundation of China under Grant No. 11871110, the National Key Research and Development Program of China under Grants No. 2016YFB0201200 and No. 2016YFB0201203, and Beijing Academy of Artificial Intelligence (BAAI).

\appendix
\section{Neural Network Structure\label{sec:NN}}
In Sec.~\ref{sec:theory} the neural network used to fit the correction energy $E_\delta$ from descriptors $\vd^I$ is denoted briefly by $\sum_I \cF^\mr{NN}\pqty{\vd^I }$. Here we provide a detailed description on the neural network structure we used. Since all atomic descriptors $\vd^I$ share the same fitting function $\cF^\mr{NN}$, we will omit the index $I$ hereafter.

The main part of $\cF^\mr{NN}$ is constructed as a feedforward neural network with optional skip connection. To speed up the training process, it also supports preprocessing and prefitting the input data on the training set. More specifically, $\cF^\mr{NN}$ has the following form:
\begin{equation}
    \cF^\mr{NN} = \cL^\mr{pre}\pqty{\bar{\vd}} + \cL^\mr{out} \circ \cL^{L} \circ \cdots \circ \cL^2 \circ \cL^1 \pqty{\bar{\vd}}
\end{equation}
where the symbol ``$\circ$'' means function composition and $\cL^p$ ($p \in \qty{1,2,\dots,L}$) denotes the mapping from layer $p-1$ to $p$ in the neural network, which we will explain later. 

The term $\cL^\mr{pre}\pqty{\bar{\vd}} = \bm{W}^\mr{pre} \cdot \bar{\vd} + \bm{b}^\mr{pre}$ corresponds to the prefitting procedure and $\bar{\vd} = \pqty{\vd - \bm\mu^\mr{pre}} / {\bm\sigma^\mr{pre}}$ is the preprocessed descriptor, where $\bm{W}^\mr{pre}$ and $\bm{b}^\mr{pre}$ can be determined through ridge regression on the training set and $\bm\mu^\mr{pre}$ and $\bm\sigma^\mr{pre}$ can be taken as the mean and standard variance of each component of the descriptors over the training set. Alternatively, setting $\bm{W}^\mr{pre}$, $\bm{b}^\mr{pre}$ and $\bm\mu^\mr{pre}$ to 0 and  $\bm\sigma^\mr{pre}$ to 1 can turn off the preprocessing completely. These behaviors are controlled by \verb|prefit|, \verb|preshift| and \verb|prescale| keywords under \verb|preprocess_args| of \verb|train_input|.

For each layer $\cL^p$ in the neural network, we have 
\begin{equation}
    \vd^p = \cL^p\pqty{\vd^{p-1}} = \varphi\pqty{\bm{W}^p \cdot \vd^{p-1} + \bm{b}^p},
\end{equation}
where $\vd^p$ are the values of neurons in layer $p = 1,2,\dots,L$ for and $M_p$ the number of neurons controlled by the keyword \verb|hidden_sizes| under \verb|model_args|. By default (and in the example above), we use three hidden layers ($L=3$) and 100 neurons for each hidden layer ($M_p=100$). In particular, $\vd^0 = \bar{\vd}$ is the preprocessed input of the neural network. The weight $\bm{W}^p \in \RR^{M_p \cross M_{p-1}}$ and bias $\bm{b}^p \in \RR^{M_p}$ are parameters to be optimized and the activation function $\varphi$ is applied component-wisely and specified by \verb|actv_fn| keyword, defaults to GELU\cite{hendrycks2016gaussian}. When the keyword \verb|use_resnet| is set to true and $M_p = M_{p-1}$, skip connection is added to the layer construction
\begin{equation}
    \vd^p = \cL^p\pqty{\vd^{p-1}} = \vd^{p-1} + \varphi\pqty{\bm{W}^p \cdot \vd^{p-1} + \bm{b}^p},
\end{equation}
to facilitate the training process.

The output layer $\cL^\mr{out}$ does not have an activation function and is just a linear transformation,
\begin{equation}
    \cL^\mr{out}\pqty{\vd^{L}} = \frac{1}{\sigma^\mr{out}}\pqty{\bm{W}^\mr{out} \cdot \vd^{L} + \bm{b}^\mr{out}},
\end{equation}
with  weight $\bm{W}^p \in \RR^{1 \cross M_L}$ and bias $\bm{b}^p \in \RR$ are parameters to be optimized, and $\sigma^\mr{out}$ a scalar factor specified by \verb|output_scale| in advance to speed up training. 

In the case of training a DeePKS model for a relatively complicated system, the sorting of eigenvalues in the construction of descriptors may lead to non-smooth derivatives on specific points, making the calculation hard to converge. In that case, we provide a smooth embedding for the descriptors $\vd^0 = \cM\pqty{\bar{\vd}}$, achieved by taking a thermal average under different ``inverse temperatures'' $\beta_k$, over the eigenvalues $\qty{d_{nlm}}$, $m=-l, \dots, l$ in the same shell specified by $n$ an $l$, 
\begin{equation}
    d^0_{nlk} = \cM_k\pqty{\qty{\bar{d}_{nlm}}} = \frac{\sum_m \bar{d}_{nlm} \exp(\beta_k \bar{d}_{nlm}) }{\sum_m \exp(\beta_k \bar{d}_{nlm}) },
\end{equation}
where $\beta_k$ are trainable parameters that by default are taken evenly between $-5$ and $5$ with number of $k$ equals to the number of $m$. The embedding can be enabled by setting \verb|embedding| to \verb|thermal| in \verb|model_args|.
\bibliographystyle{elsarticle-num}
\bibliography{ref_abbr}

\end{document}